\documentstyle[prc,preprint,tighten,aps]{revtex} 

\preprint{\vbox{. \hfill UM-P-96/49, RCHEP 96/05}} 

\title{Dirac-Foldy term and the electromagnetic polarizability of the
neutron}

\author{M.\ Bawin\thanks{Electronic-mail address:U216204@vm1.ulg.ac.be}}
\address{Universit\'{e} de Li\`{e}ge, Institut de Physique B5, Sart Tilman,
4000
Li\`{e}ge 1, Belgium }

\author{S.\ A.\ Coon\thanks{Electronic-mail address: coon@nmsu.edu}
\thanks{Permanent address: Physics Department, New Mexico State University,
 Las
Cruces, NM 88003, USA}}
\address{ School of Physics, Research Center for High Energy Physics, 
University of Melbourne, Parkville, Victoria, Australia, 3052}

\begin{document}

\maketitle

\begin{abstract} We reconsider the Dirac-Foldy contribution $\mu^2/m$ to
the
neutron electric polarizability. Using a Dirac equation approach to
neutron-nucleus scattering, we review the definitions of Compton continuum
($\bar{\alpha}$), classical static ($\alpha^n_E$), and Schr\"{o}dinger
($\alpha_{Sch}$) polarizabilities and discuss in some detail their
relationship.
The latter $\alpha_{Sch}$ is the value of the neutron electric
polarizability as
obtained from an analysis using the Schr\"{o}dinger equation. We find in
particular $ \alpha_{Sch} = \bar{\alpha} - \mu^2/m$ , where $\mu$ is the
magnitude of the magnetic moment of a neutron of mass $m$. However, we
argue
that the static polarizability $\alpha^n_E$ is correctly defined in the
rest
frame of the particle, leading to the conclusion that twice the Dirac-Foldy
contribution should be added to $\alpha_{Sch}$ to obtain the static
polarizability $\alpha^n_E$.
\end{abstract}
\vskip20pt

\baselineskip 0.6cm	

\newpage

\section{Introduction}

The electromagnetic polarizabilities of the nucleon continue to attract
interest
because of their importance for the understanding of the substructure of
the
nucleon. The proton and neutron form an isospin doublet with (presumably)
similar substructure, so it is expected that comparing experimental
polarizabilities of the two would lead to more insight into this
substructure~\cite{NH,chiral}. This comparison should take into account the
different {\em definitions} of electromagnetic polarizabilities actually
used in
the measurements on the proton and the neutron. The electric and magnetic
polarizabilities of the proton can be defined\cite{JS,Friar1} and
measured\cite{pexp} via Compton scattering at relatively low energies
because of
the interference of the Rayleigh amplitude (from the polarizabilities) with
the
Thomson amplitude (from the charged proton). Because the neutron is neutral
there is no such interference and the cross section for elastic Compton
scattering is much smaller. Furthermore, the data must come from a neutron
bound
in a nucleus, say a deuteron\cite{Lucas}, and it is a challenge to
interpret it
in terms of neutron polarizabilities\cite{Wilbois,Levchuk}. An alternative
would
be to determine neutron polarizabilities via quasi-free Compton scattering,
but
the first experiment could only obtain an upper limit for the electric
polarizability~\cite{Rose}. It is expected to be redone at SAL with a
considerable reduction in the statistical error~\cite{Skopik}. The best
determination of the electric polarizability of the neutron is obtained, at
present, from low energy neutron-atom scattering\cite{Jorg}. The intense
electric field near the surface of the nucleus $^{208}$Pb induces a dipole
moment in the neutron which makes a tiny but extractable contribution to
the
scattering amplitude. The electric polarizability of the neutron is defined
as
the coefficient of the $r^{-4}$ nonrelativistic potential acting between
these
two systems. 

We wish to reexamine the relationship between the definitions of neutron
electric polarizability in use, noting that the Compton scattering definition is
manifestly relativistic and the neutron-atom scattering definition is not.
Furthermore, the Compton scattering definition actually used to extract the
electric polarizability of the proton (soon to be extended to the neutron)
does
not employ a Hamiltonian nor a wave equation and the neutron-atom
scattering
definition is in the context of the Schr\"{o}dinger equation with its
implied
Hamiltonian. The common meeting ground of these seemingly disparate
definitions
in current use is given by the relationship of each one to the classical
definition of $\alpha^n_E$ as the coefficient of ${\bf E}^2$ in $V^{pol} =
-{\textstyle\frac{1}{2}} \alpha^n_E {\bf E}^2 \approx -Q^2 \alpha^n_E/(2
r^4)$
where $V^{pol}$ represents the interaction of a neutral particle {\em at
rest}
with the Coulomb field ${\bf E}\sim Q \hat{r}/r^2$ of an infinitely heavy
charged system~\cite{IZ}. In the following we establish these
relationships.
That is, we: 

\begin{itemize}
\item[1)] remind the reader of the definition of $\alpha_s$ extracted by
experimentalists from the spin-averaged Compton cross section and the
definition
of the more intuitive Compton polarizability $\bar{\alpha}$ of, for
example,
chiral perturbation theory calculations (only the latter $\bar{\alpha}$
corresponds to a true ``deformation" effect on the nucleon.) 

\item[2)] quote the classical limit $\alpha^n_E$ of the Compton
polarizability
of a neutral particle, $\alpha^n_E = \bar{\alpha} + \frac{\mu^2}{m} =
\alpha_s
+\frac{2\mu^2}{m}$, where $\mu$ is the anomalous (in this case, total)
magnetic
moment of the neutron \cite{units}.

\item[3)] embed the Compton defined $\bar{\alpha}$ in a relativistic Dirac
description of neutron-atom scattering to establish the non-relativistic
classical limit $\alpha^n_E = \bar{\alpha} + \mu^2/m $, where the static
polarizability $\alpha^n_E$ is the coefficient of ${\bf E}^2$ in the
neutron's
rest frame. With the {\em correct} rest frame wave equation this result is
identical with the classical limit of the Compton result of 2).

\item[4)] assert that the rest frame of a neutron in an external electric
field
is defined by a vanishing value of the velocity operator, as confirmed by
experimental measurements of the Aharonov-Casher effect. 

\item[5)] note that Schmiedmayer et al. \cite{Jorg} and
others\cite{Koester} use
the Schr\"{o}dinger equation in order to analyze low energy neutron-atom
scattering experiments. The coefficient of ${\bf E}^2$ in this equation is
then
$\alpha_{Sch}$, and was considered the electric polarizability of the
neutron
from those experiments. 

\item[6)] show that the $\alpha_{Sch}$ of Schmiedmayer et al. \cite{Jorg}
and
others\cite{Koester} is neither the Compton defined $\bar{\alpha}$ nor the
static $\alpha^n_E$, but $\alpha_{Sch} = \alpha^n_E - 2\mu^2/m$.

\end{itemize}

We conclude from this chain of arguments that twice the Dirac-Foldy
contribution
$\mu^2/m$ should be added to $\alpha_{Sch}$ to obtain the static
polarizability
$\alpha^n_E$ from the existing analysis of neutron-atom scattering
experiments.
That is the message of our paper. 

Details of the discussion of Compton-defined polarizabilities are in
Section II,
and our Dirac equation discussion of the electromagnetic aspects of
neutron-atom
scattering is in Section III.

neutron-atom scattering situation, and 
controversy
about the interpretation of 
nonrelativistic 
of a
Dirac 

\section{polarizabilities in Compton scattering} Already in the earliest
experimental studies of low-energy Compton scattering from the
proton~\cite{Goldanskii} it was realized that the ``polarizabilities"
entering
into the Rayleigh amplitude had two contributions. That is, the external
electromagnetic fields both deform the particle and act upon the static
distribution of the electric charge and the magnetic moment. Thus we read
in
Ref. \cite{Goldanskii}: ``the term `polarizability' used here is not
equivalent
to the one normally employed for neutral particles". The situation is made
more
difficult by the fact that the nucleon is a spin $1/2$ particle with an
anomalous magnetic moment and is described by the Dirac equation. The
Compton
scattering matrix for a spin $1/2$ particle is the sum of six Lorentz
invariant
quantum field amplitudes which are free of kinematic singularities and
constraints \cite{JS}. These six amplitudes each contain single nucleon
pole
terms (a structureless Dirac nucleon with charge Z and magnetic moment
$\mbox{\boldmath $\mu$}$ and on-shell vertices). For each amplitude the
remainder is called a continuum contribution and is now free of both
kinematic
singularities and dynamical singularities (from the nucleon poles). If one
thinks of polarizabilities as a ``deformation" effect on the structure of
the
nucleon they would seem to be most naturally defined in terms of the latter
continuum contributions. However, there is a freedom in the definition of
Compton polarizabilities of spin $1/2$ particles due to the fact that the
entire
Compton matrix is not measured. Instead present experiments measure the
spin
averaged cross section which corresponds to only the spin independent part
of
the Compton matrix. Bernab\'{e}u and Tarrach \cite{BT} (BT) note that the
nucleon pole contributions to the amplitudes of the complete spin
$\frac{1}{2}$
Compton scattering matrix generate both pole {\em and} continuum
contributions
to the spin averaged amplitudes actually measured as a differential
cross-section. The most common choice (labeled $\alpha_s$ by BT and used in
this
paper as well) includes in the ``polarizability" terms from the magnetic
moment
of the structureless Dirac particle. For this choice the differential cross
section takes the form \begin{equation}
\left({d\sigma\over d\Omega}\right)_L = \left({d\sigma\over
d\Omega}\right)^{\rm
proton}_{\rm poles} - {\alpha\over m_p} \omega^2 \left[1-{3\omega\over
m_p}(1-z)\right]\left[(1+z^2) \alpha^p_s + 2z \beta^p_s\right] + {\cal
O}(\omega^4)\ \ \ , \label{Compt} \end{equation} valid through the first
three
moments of the photon lab energy $\omega$, where $z =
\cos\theta_L$\cite{JS,units}. The Born terms of the invariant amplitudes go
into
the Thomson cross section for a pointlike particle with mass $m$ and charge
Z in
its rest frame and the (actually used) Powell cross section of
(\ref{Compt}),
also for a pointlike particle, but one which includes an anomalous magnetic
moment~\cite{Powell}. Equation (\ref{Compt}) or its extension to higher
energy
is used to extract $\alpha^p_s$ from proton Compton scattering data
\cite{pexp}.

Now we return to the classical definition of $\alpha^n_E$ as the
coefficient of
an ${\bf E}^2$ or $r^{-4}$ term in a nonrelativistic wave equation. The
concept
of a potential as it applies to the interaction of two systems in
relativistic
quantum field theory and the computation of such Van der Waals potentials
due to
induced dipoles (when the systems are far apart) has been discussed
extensively
by Feinberg and Sucher~\cite{FS}. The electromagnetic forces between
charged
and/or neutral systems are due to the exchange of photons and can be
calculated
with the aid of dispersion relations from the relativistic Compton
amplitudes of
photons scattering from the system. Specifically, the potential is to be
defined
iteratively in such a way that when used in a specified two-body (Dirac)
wave
equation in the c.m system it will reproduce, up to a given order, the
field-theory amplitude associated with one-photon- and two-photon-exchange
graphs. The potential then can be reduced to the Schrodinger form and its
long-ranged part compared with the nonrelativistic polarizability
potential.
Thus there is a clear line of connection between the electric Compton
polarizability and the classical electric polarizability which does not
depend
upon an intuitively appealing but theoretically uncertain mixture of
relativistic and non-relativistic concepts~\cite{Shekhter}.

This program of connecting classical polarizability with the low energy
Compton
scattering parameters has been carried out by Feinberg and Sucher for a
variety
of systems (two spinless and uncharged particles, one neutral spinless and
one
charged spin-${\textstyle\frac{1}{2}}$ particle ~\cite{FS1992}, etc.), all
but
the one relevant to our examination of neutron-atom scattering. The
long-range
potential of these two systems, a very massive charged spin zero nucleus
and a
neutral spin-${\textstyle\frac{1}{2}}$ neutron (with an anomalous magnetic
moment) has been worked out by Bernab\'{e}u and Tarrach\cite{BT}. They note
that
the nucleon pole contributions to the six amplitudes of the complete spin
${\textstyle\frac{1}{2}}$ Compton scattering matrix generate both pole {\em
and}
continuum contributions to the spin averaged amplitudes actually measured
as a
differential cross-section. Thus one can define (in their notation but our
units~\cite{units}) an $\alpha_s$ which does include a term with the
anomalous
magnetic moment ($-(e\mu Z/m^2 + \mu^2/m)$) or a $\bar{\alpha}$ which is
given
only in terms of the continuum (non-pole) contributions of the spin
averaged
amplitudes. The former definition corresponds to the actual analysis of
Compton
scattering data~\cite{pexp} according to (\ref{Compt}) and the latter is
advocated by Bernab\'{e}u and Tarrach and used in some theoretical
treatments~\cite{chiralCompt}. The latter polarizabilities so defined do
not
receive any contribution from Born graphs involving the anomalous magnetic
moment of the proton and the neutron. The polarizabilities are entirely
given in
terms of the continuum part of the Compton amplitude. Equivalently
$\bar{\alpha}$ is {\em defined} to be zero for a point neutral Dirac
particle.
In the classical limit of a static electric field acting on a neutral
particle
of mass $m$ and magnetic moment $\mu$ the coefficient of the $r^{-4}$
potential
which survives is given by
\begin{equation} \alpha^n_E = \bar{\alpha} + \frac{\mu^2}{m} = \alpha_s
+\frac{2\mu^2}{m}\,\,. \label{BT} \end{equation} It is then this sum which
is
measured in the scattering of neutrons by heavy nuclei at low energies.

\section{Dirac equation analysis of neutron-atom scattering} 

We see then how a natural definition of the polarizability of a neutral
particle
$\bar{\alpha}$ arises in the context of Compton scattering and understand
its
connection via the Feinberg-Sucher-Bernab\'{e}u-Tarrach analysis with the
polarizability potential $V^{pol} = -{\textstyle\frac{1}{2}} 
\alpha^n_E{\bf E}^2$ of a Schr\"{o}dinger analysis of low energy neutron
scattering. We now establish such a connection again, this time starting
from a
relativistic Dirac description of the neutron-nucleus scattering. We derive
the
neutron polarizability as the nonrelativistic limit of a relativistic Dirac
Hamiltonian: 

\begin{equation} H^D = \beta m + \mbox{\boldmath $\alpha$} \cdot {\bf p} -
i \mu
\beta \mbox{\boldmath $\alpha$} \cdot {\bf E} - {\textstyle\frac{1}{2}}
\bar{\alpha} {\bf E}^2 \,\,\, , \label{FoldyHam}
\end{equation} where the first three terms comprise the standard formula
\cite{Powell,Foldy,Obermair} for a point neutral Dirac particle with an
anomalous magnetic moment $\mbox{\boldmath $\mu$}$  in an electric field.
As in
the BT treatment of Compton scattering, $\bar{\alpha}$ is that part of the
neutron's polarizability that does not contain the nucleon magnetic moment
$\mu$. Even so, the nonrelativistic reduction of (\ref{FoldyHam}) has a
term in
${\bf E}^2 $ in addition to the nominal polarizability $\bar{\alpha}$:

\begin{equation} (\frac{{\bf p}^2}{2m} - \frac{{\bf p}}{m}\cdot ({\bf E}
\times
\mbox{\boldmath $\mu$} ) + \frac{\mu}{2m}(\mbox{\boldmath $\nabla$}\cdot
{\bf
E}) + \frac{\mu^2 {\bf E}^2}{2m} - {\textstyle\frac{1}{2}} \bar{\alpha}
{\bf
E}^2)\psi = E\psi\,\,\, ,
\label{nonrel}
\end{equation} where we have neglected interaction terms that vanish faster
then
$r^{-4}$ at large distance. The second and third terms in (\ref{nonrel}) are the
Schwinger term arising from the interaction between the (moving) magnetic
moment
of the neutron and the electric field of the atom, and the  Foldy-Darwin
scattering from the electric charge distribution of the atom (nucleus +
electrons) These terms are taken into account in the nonrelativistic
analysis of
neutron-atom scattering \cite{Leeb,Schthesis,Amos}. Then it would seem that
the
coefficient in the polarizability potential is \begin{equation}
	\alpha_{Sch} = \bar{\alpha} - \mu^2/m	\label{strawman}
\end{equation} rather than the Compton defined $\bar{\alpha}$. 

This (premature) result could have been anticipated by Foldy's observation
that a
structureless (point) neutral Dirac particle with an anomalous magnetic
moment
$\mu$ in a homogeneous static electric field $\vec{E}$ is an exactly
soluble
model~\cite{Foldy}. That is, \begin{equation} H = \beta m + \mbox{\boldmath
$\alpha$} \cdot {\bf p}  - i \mu \beta \mbox{\boldmath $\alpha$} \cdot \bf
E
\,\,\,. \label{oldFoldyHam}
\end{equation} This Hamiltonian can be diagonalized, in the frame where
${\bf p}
= 0$, by simply squaring and one finds the energy eigenvalues $ W = \pm
\sqrt{m^2 + \mu^2 {\bf E}^2} \simeq \pm [ m + \mu^2 {\bf E}^2/2m + \cdots]
$.
The nonrelativistic limit of this model has a positive coefficient of $
{\bf
E}^2$ which implies a {\em negative} polarizability of magnitude $\mu^2/m $
from
this Dirac-Foldy term, just as we find in (\ref{strawman}). Moreover, now
we see
that the analysis leading to (\ref{strawman}) has been carried out in the
frame
${\bf p} = 0$.

But we must be careful to define polarizability of a particle in that
particle's
rest frame \cite{IZ} and the rest frame of a particle is defined by a
vanishing
value of the velocity operator $\bf v$. The {\em particle} velocity
operator is
given as a derivative of the Hamiltonian on the left hand side of
(\ref{nonrel}):
\begin{equation} {\bf v} \equiv \frac{\partial H}{\partial {\bf p}} =
\frac{1}{m} [ {\bf p} - ({\bf E} \times \mbox{\boldmath $\mu$} )]\,\,. 		
\label{velocity}
\end{equation} That is, the (${\bf v}= 0)$ frame is {\em not} the (${\bf
p}= 0)$
frame leading to (\ref{strawman}). From (\ref{velocity}) one can rewrite
(\ref{nonrel}) in the form familiar from discussions of the Aharonov-Casher
effect~\cite{AC,Hagen,He}:
\begin{equation} [\frac{1}{2m} (({\bf p} - ({\bf E} \times \mbox{\boldmath
$\mu$} ))^2 + 
\frac{\mu}{2m}(\mbox{\boldmath $\nabla$}\cdot {\bf E}) - \frac{\mu^2 {\bf
E}^2}{2m} - {\textstyle\frac{1}{2}} \bar{\alpha} {\bf E}^2)]\psi = E\psi
\label{nonrelac}
\end{equation} From this equation one identifies
\begin{equation}
	\alpha^n_E = \bar{\alpha} + \mu^2/m	\label{result}
\end{equation} to be the coefficient of ${\bf E}^2$ in the particles rest
frame
(${\bf v}= 0)$, and $\alpha^n_E$ is then the static polarizability of the
neutron.

This rest frame result is in agreement with the Compton scattering analysis
of BT
in Eq. (\ref{BT}). In order to avoid any possible misunderstanding, let us
emphasize that our discussion of polarizability terms in neutron-nucleus
scattering is entirely in the framework of the nonrelativistic limit of the
Dirac equation. From that viewpoint, one may argue that it provides an
intuitive way of understanding the results of Bernabeu and Tarrach
\cite{BT}
which were obtained from dispersion relations calculations. We do, however,
discuss in detail the form of the nonrelativistic wave equation
((\ref{nonrelac}) rather than (\ref{nonrel})) to be used in conjunction with
the BT results.\par
The observation~\cite{Opat} of the phase shift predicted by
Aharonov-Casher\cite{AC} for a neutral particle with a magnetic moment
(neutron)
diffracted around a line of electric charge shows conclusively that
(\ref{nonrelac}) is the  correct rest frame equation. For a neutron
diffracting
around a line charge in a region where
$\mbox{\boldmath $\nabla$}\cdot {\bf E}=0$,  the Aharonov-Casher phase
shift is
obtained by evaluating the line integral of ${\bf p} = m {\bf v} + ({\bf E}
\times \mbox{\boldmath $\mu$} )$ along the path of the diffracted neutrons. (Of
course, $\bar{\alpha}$ could not play any role in this macroscopic
experiment,
and the fact that the term ${\textstyle\frac{\mu^2 {\bf E}^2}{2m}}$
disappears
in the Aharonov-Casher geometry is explained in Refs. \cite{Hagen,He}) More
recent experiments involving neutral atoms with magnetic moments have
measured
Aharonov-Casher phase shifts to within a few per cent of the theoretically
predicted value\cite{ACatom}.

The neutron optics experiment, fortified by more exact measurements with
atomic
systems, demonstrates the Aharonov-Casher insight that velocity is the
meaningful relativistic kinematic operator for a neutron in an external
electric
field. We have used this insight to define the correct static
polarizability of
a neutral particle with a magnetic moment. From Eqs. (\ref{strawman}) and
(\ref{result}) it is clear that the $\alpha_{Sch}$ measured in the
experiments
of Schmiedmayer et al \cite{Jorg} and others \cite{Koester} is neither
$\bar{\alpha}$ nor $\alpha^n_E$. Indeed from (\ref{strawman}) and
(\ref{result})
we learn that \begin{equation}
	\alpha_{Sch} = \alpha^n_E - 2\mu^2/m	\label{bigwhoop}
\end{equation} Numerically,
\begin{equation}
	|\alpha_{Sch} - \alpha^n_E| = 1.2 \times 10^{-4}\,\,\,{\rm fm}^3 			
\label{numbwh}
\end{equation} can be compared with
\begin{eqnarray}
	\alpha_{Sch} (\cite{Jorg}) & = & 12 \pm 1.5 \pm 2.0 \times
10^{-4}\,\,\,{\rm
fm}^3 \nonumber \\
	\alpha_{Sch}(\cite{Koester}) & = & 0.0 \pm 5 \times 10^{-4}\,\,\,{\rm
fm}^3
\label{expdis}
\end{eqnarray} This difference is about 10\% on the scale of the
Schmiedmayer et
al. result\cite{Jorg} and quite significant for the central value of the
Koester
et al. result\cite{Koester}. Both results came from a Schr\"{o}dinger
equation
analysis like Eq. (\ref{nonrel}). The discrepancy in Eq. (\ref{expdis})
perhaps
comes from the treatment of individual terms in the electromagnetic
interaction
of Eq.\ (\ref{nonrel}) or from the treatment of the strong interaction
between
the neutron and the nucleus. In any case, our Dirac equation analysis has
nothing to say about the origin of the present experimental discrepancy. We
note
that these experiments are being repeated \cite{Kopecky,Leebpri} with an
expected experimental error smaller in magnitude than our correction term
of Eq.
(\ref{numbwh}).

Finally we note that L'vov\cite{Lvov} obtains (by another argument) a
relationship between $\alpha^n_E$ and $\bar{\alpha}$ which agrees with
(\ref{strawman}) if one equates $\alpha^n_E$ and $\alpha_{Sch}$ as he does.
In
the neutron rest frame, however, the correct relationship is that of Eq.
(\ref{bigwhoop}).

In summary, we have reviewed the definition of the electrical
polarizability of
a neutral spin ${\textstyle\frac{1}{2}}$ particle with a magnetic moment
$\mu$
in the analysis of Compton scattering. We have shown how a Dirac equation
analysis of low energy neutron-atom scattering, yields a static
polarizability
{\em defined in the rest frame of the neutron}. Our result (\ref{bigwhoop})
means that twice the Dirac-Foldy contribution $\mu^2/m$ should be added to
the
existing Schr\"{o}dinger values to obtain the static polarizability of the
neutron.

\acknowledgments SAC is grateful for the hospitality of the University of
Melbourne and acknowledges support in part by NSF grant PHY-9408137. The
work of
MB  was supported by the National Fund for Scientific Research, Belgium. We
are
grateful to Helmut Leeb for discussions and for sending us unpublished
material
on the Schr\"{o}dinger analysis of neutron-atom scattering. SAC
acknowledges
useful discussions with Tony Klein, Bruce McKellar, Mike Scadron, and Andy
Rawlinson.


\begin{thebibliography}{99}
\bibitem{units} We use Gaussian units so that the fine structure constant
$\alpha= e^2/\hbar c \approx 1/137$. The polarizabilities then are of order
$10^{-4}$ fm$^3$. The magnetic moment is the true magnetic moment $\mu = (Z
+
\kappa)\mu_N = \kappa (e \hbar/2mc)$ where $Z=0$ and $\kappa = -1.91$ for
the
neutron. The nuclear Bohr magneton $\mu_N\sim 5 \times 10^{-14} {\rm MeV
T}^{-1}$ is a positive number. We set $\hbar=c=1$ in the formulas of the
text.
\end{thebibliography}
\end{document}